\documentclass{ws-procs9x6-cpt22}
\begin{document}

\newcommand{\refeq}[1]{(\ref{#1})}
\def\etal {{\it et al.}}

\title{Lorentz Violation in Electromagnetic Moments of Fermions}

\author{Javier Monta\~no-Dom\'inguez,$^1$ H\'ector Novales-S\'anchez,$^2$ M\'onica Salinas,$^2$, and J. Jes\'us Toscano$^2$}

\address{$^1$Facultad de Ciencias F\'isico Matem\'aticas,\\
Universidad Michoacana de San Nicol\'as de Hidalgo,
Morelia, Michoac\'an, Mexico.}

\address{$^2$Facultad de Ciencias F\'isico Matem\'aticas,\\ Benem\'erita Universidad Aut\'onoma de Puebla,
Puebla, Puebla, Mexico}

\begin{abstract}
Lorentz-violating Yukawa couplings lying within the renormalizable part of the SME generate Lorentz-invariant one-loop contributions to electromagnetic moments of fermions. This note provides a discussion of such contributions and presents bounds on SME coefficients from experimental data on electromagnetic moments. Constraints as restrictive as $\sim10^{-14}$ are found.
\end{abstract}

\bodymatter

\section{Lorentz-violating Yukawa interactions}
This paper is based on Refs.~[\refcite{AMNST}], where a comprehensive discussion and further details can be found.
The Yukawa sector of the minimal SME (mSME) reads\cite{CoKo}
\begin{eqnarray}
{\cal L}^{\rm mSME}_{\rm Y}=&&-\frac{1}{2}(H_U)^{AB}_{\mu\nu}\overline{Q}_A\tilde{\phi}\sigma^{\mu\nu}U_B-\frac{1}{2}(H_D)^{AB}_{\mu\nu}\overline{Q}_A\phi\sigma^{\mu\nu}D_B
\nonumber \\ &&
-\frac{1}{2}(H_L)^{AB}_{\mu\nu}\overline{L}_A\phi\sigma^{\mu\nu}R_B+{\rm H.c.},
\end{eqnarray}
with $\tilde{\phi}=i\sigma^2\phi^*$ and $\phi$ the SM Higgs doublet, $Q_A$ an ${\rm SU}(2)_L$ quark doublet, $L_A$ an ${\rm SU}(2)_L$ lepton doublet, $U_B$ and $D_B$ up- and down-type ${\rm SU}(2)_L$ quark singlets, respectively, and $R_B$ a charged-lepton singlet of ${\rm SU}(2)_L$. Capital-letter indices denote fermion flavors, whereas Greek indices label spacetime coordinates. The Lorentz-violation coefficients in ${\cal L}_{\rm Y}^{\rm mSME}$ are $(H_U)_{\mu\nu}^{AB}$, $(H_D)_{\mu\nu}^{AB}$, and $(H_L)_{\mu\nu}^{AB}$, all of them bearing both spacetime and fermion-flavor indices. After spontaneous symmetry breaking, coefficients $(Y_f)_{\mu\nu}=U^{f\dag}_L(H_f)_{\mu\nu}U^f_R$, are defined, with $U_L^f$ and $U_R^f$ the unitary matrices defining the fermion mass-eigenspinor basis in the SM. The resulting Lagrangian thus acquires the form
\begin{equation}
{\cal L}^{\rm mSME}_{\rm Y}=-\frac{1}{2}(v+H)\sum_{f=l,u,d}\overline{f}_A\big[ (Y_f)_{\mu\nu}^{AB}P_L+(Y_f)_{\mu\nu}^{BA*}P_R \big]\sigma^{\mu\nu}f_B,
\label{LSMEY2}
\end{equation}
with $H$ the Higgs field, and $P_L$, $P_R$ the chiral projectors. Since $(Y_f)_{\mu\nu}=-(Y_f)_{\nu\mu}$, which characterizes the electromagnetic tensor $F_{\mu\nu}$ as well, we define the 3-vector background fields ${\bf e}_f^{AB}$ and ${\bf b}_f^{AB}$, for some fermion $f_A$, by $(Y_f)^{AB}_{0j}=({\bf e}^{AB}_f)_j$ and $(Y_f)^{AB}_{jm}=\epsilon_{jmk}({\bf b}^{AB}_f)^{k}$. These are the mSME parameters to be compared with experimental data.

Feynman rules from Eq.~(\ref{LSMEY2}) yield one-loop diagrams contributing to the electromagnetic moments (EMM) of fermions at second order in SME insertions. The large set of such diagrams is provided in Refs.~[\refcite{AMNST}]. By summing all the diagrams together, we find a $\gamma f_Af_A$ vertex-function contribution
\begin{equation}
\begin{gathered}
\vspace{0.15cm}
\includegraphics[width=1.8cm]{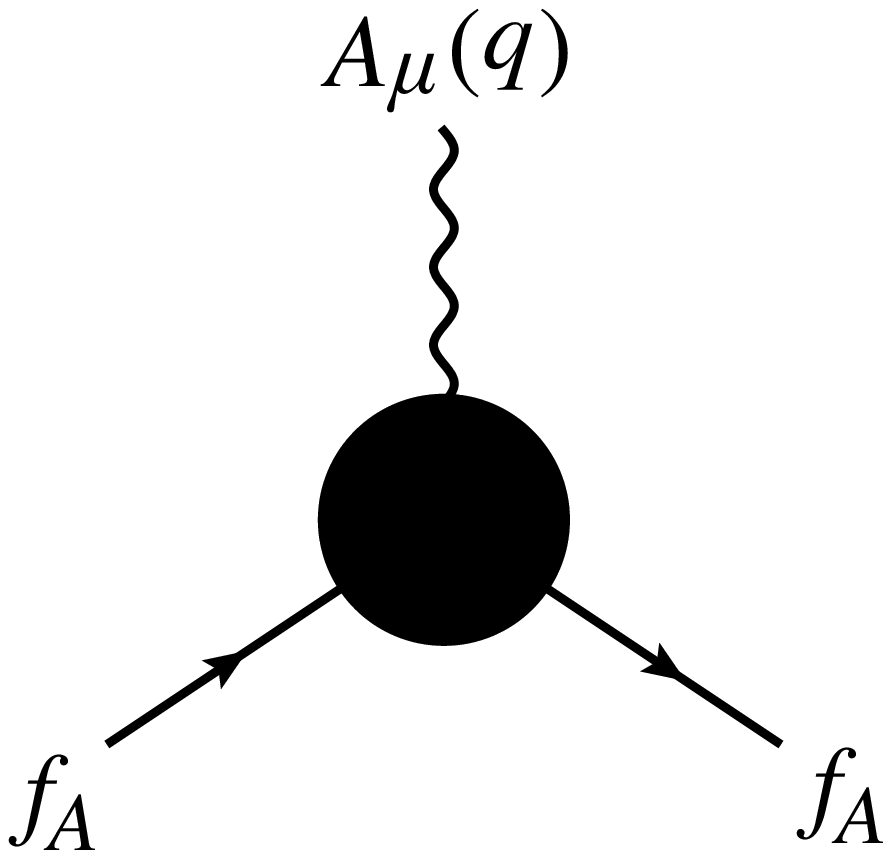}
\end{gathered}
=
\overline{{\cal U}}_A\Big[ F_A^{\rm M}(q^2)\sigma_{\mu\nu}q^\nu+F^{\rm E}_A(q^2)\sigma_{\mu\nu}q^\nu\gamma_5+\cdots \Big]{\cal U}_A.
\label{ffApar}
\end{equation}
Here, ${\cal U}_A$ is the momentum-space spinor associated with the external fermion $f_A$. Moreover, $q$ is the incoming momentum of the external photon, which we assume to be off shell. Later, the on-shell condition $q^2=0$ is taken in order to define the mSME Yukawa-sector contributions to the anomalous magnetic moment (AMM), $a_A^{\rm SME}=F_{A}^{\rm M}(q^2=0)$, and to the electric dipole moment (EDM), $d^{\rm SME}_A=F^{\rm E}_A(q^2=0)$, of the fermion $f_A$. Note that the magnetic and electric form factors, $F_A^{\rm M}(q^2)$ and $F_A^{\rm E}(q^2)$, preserve symmetry under both observer and particle Lorentz transformations. The ellipsis in Eq.~(\ref{ffApar}) represents other terms, involving various form factors, most of them violating Lorentz invariance. In the present paper contributions to diagonal EMMs are exclusively considered. While flavor changes in the mSME Yukawa sector induce contributions to transition moments, such a calculation lies beyond the scope of the present discussion. 

Following the tensor-reduction method,\cite{PaVe} we executed an intricate calculation involving a plethora of contributing diagrams, in which either photons, or $W$ bosons, or $Z$ bosons, or Higgs bosons participate through virtual loop lines. Since some of the contributing diagrams are gauge dependent, a gauge choice was mandatory, so we used unitary gauge, thus reducing the number of diagrams. In this framework, we made sure that the amplitude fulfills the Ward identity, with respect to the electromagnetic field. While a drawback of our gauge choice is a latent complication in the elimination of ultraviolet divergences from the AMM and EDM contributions, our results turned out to be finite in this sense.

A diagrammatic expression of the mSME Yukawa-sector contribution to $\gamma f_Af_A$, previously displayed in Eq.~(\ref{ffApar}), is
\begin{eqnarray}
\begin{gathered}
\vspace{0.15cm}
\includegraphics[width=1.8cm]{ffA}
\end{gathered}
=
& \displaystyle
\sum_B\big(
\begin{gathered}
\vspace{-0.055cm}
\includegraphics[width=1.16cm]{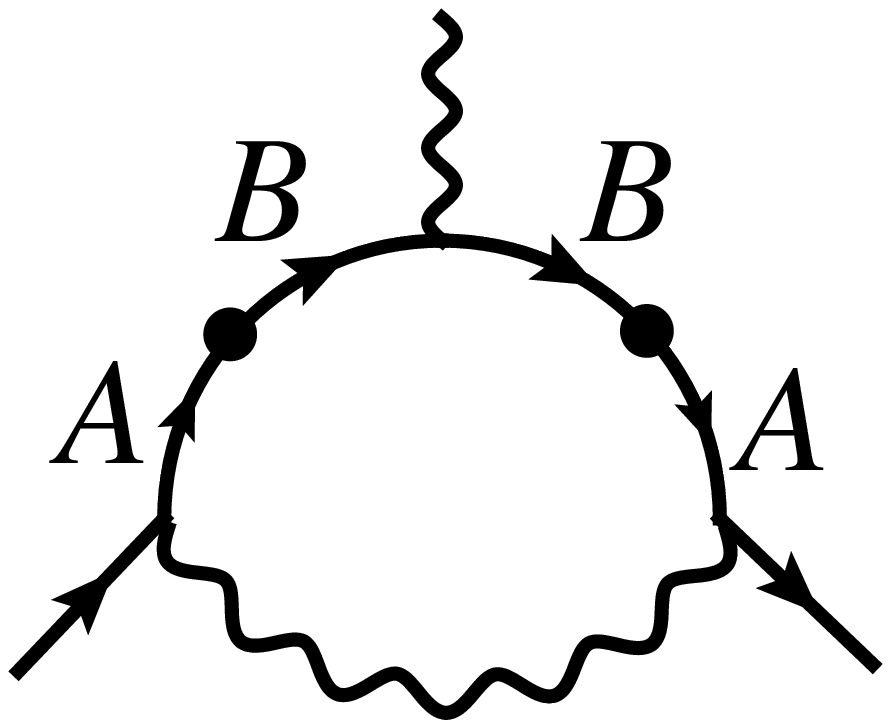}
\end{gathered}
+
\begin{gathered}
\vspace{-0.055cm}
\includegraphics[width=1.16cm]{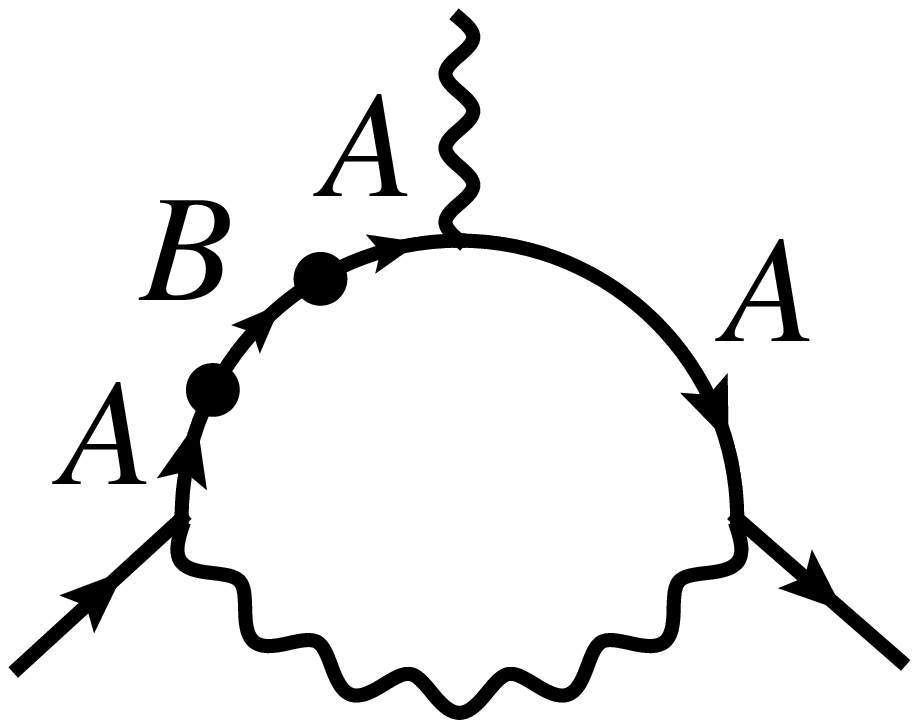}
\end{gathered}
+
\begin{gathered}
\vspace{-0.055cm}
\includegraphics[width=1.16cm]{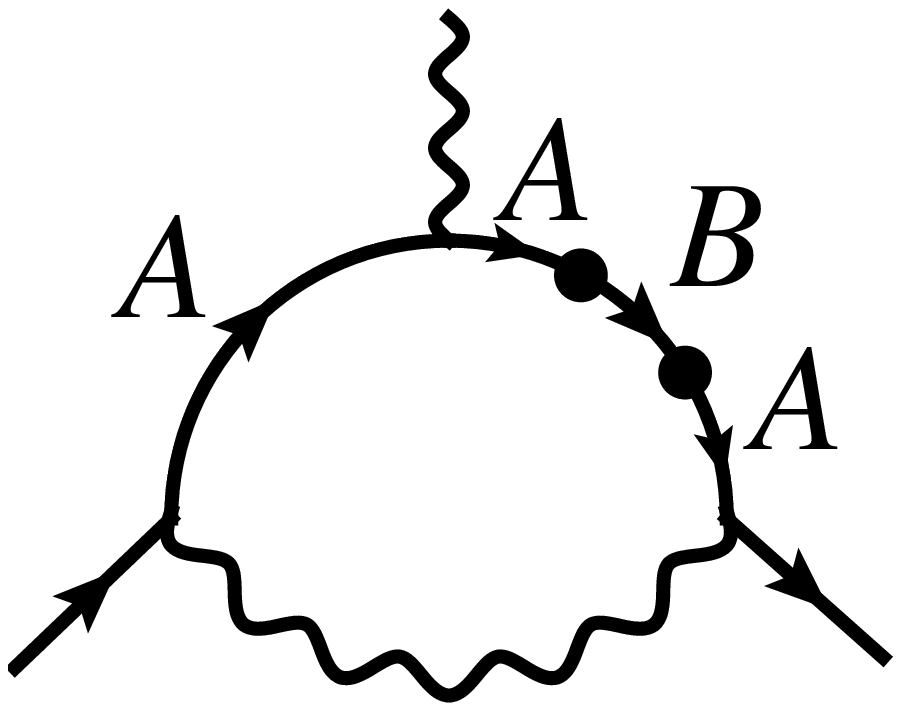}
\label{Aqqstructure}
\end{gathered}
\big)+\,\cdots,
\end{eqnarray}
where those diagrams explicitly shown, with the internal loop line corresponding to a photon, are the ones producing the dominant contributions to EMMs. From such diagrams, flavor changes induced by SME two-point insertions can be appreciated. This contribution includes a sum over fermion flavors $B$. The second and third diagrams carry infrared divergences, which do not vanish from the amplitude, thus meaning that these quantities are not observables. As discussed in detail in Refs.~[\refcite{AMNST}], cancellation of such divergences is expected to happen at the cross-section level through Bremsstrahlung diagrams, with the assumption of final-state soft-photon emission. Taking this for granted, we find finite contributions to the electromagnetic factors, given by
\begin{eqnarray}
\displaystyle
a^{\rm SME}_{f,A}=&&
\displaystyle
\sum_B\big[ \tilde{a}_{f,AB}\big( |{\rm Re}\,{\bf e}_f^{AB}|^2+|{\rm Re}\,{\bf b}_f^{AB}|^2 \big)
\nonumber \\&& \displaystyle
+\hat{a}_{f,AB}\big( |{\rm Im}\,{\bf e}_f^{AB}|^2+|{\rm Im}\,{\bf b}_f^{AB}|^2 \big) \big],
\\ \nonumber \\
\displaystyle
d^{\rm SME}_{f,A}=&&
\displaystyle
\sum_B\tilde{d}_{f,AB}\big( |{\rm Re}\,{\bf e}_f^{AB}||{\rm Im}\,{\bf b}_f^{AB}|+|{\rm Re}\,{\bf b}_f^{AB}||{\rm Im}\,{\bf e}_f^{AB}| \big).
\end{eqnarray}
In these equations, the coefficients $\tilde{a}_{f,AB}$, $\hat{a}_{f,AB}$, and $\tilde{d}_{f,AB}$ are functions of the fermion masses. Sums over fermion flavors $\sum_B$ take into account all possible virtual-fermion lines in contributing diagrams. 

Next, we consider the current best measurement of the proton magnetic moment, reported in Ref.~[\refcite{protonmm}] to be $\mu_p=2.7928473446(8)\mu_N$, with $\mu_N$ the nuclear magneton. We use the most stringent bound on the neutron EDM as well, which, according to Ref.~[\refcite{neutronedm}], is $|d_n|<1.8\times10^{-26}e\cdot{\rm cm}$. To connect the contributions from individual up and down quarks to nucleon EMMs, we used $a^{\rm SME}_p=a^{\rm SME}_u4/3-a_d^{\rm SME}/3$ and $d_n^{\rm SME}=d_d^{\rm SME}4/3-d_u^{\rm SME}/3$. From the error in the measurement of $\mu_p$, a set of bounds within $10^{-7}$ to $10^{-11}$ is achieved.\cite{AMNST} The most restrictive constraints are set on $|{\rm Re}\,{\bf e}_u^{ut}|$, $|{\rm Re}\,{\bf b}_u^{ut}|$, $|{\rm Im}\,{\bf e}_u^{ut}|$, $|{\rm Im}\,{\bf b}_u^{ut}|$, all restricted to be $<7.156\times10^{-11}$. The neutron EDM experimental bound is used to further bound SME coefficients. Such restrictions, given in Ref.~[\refcite{AMNST}], lie within $10^{-9}$ to $10^{-12}$, with the most restrictive limits corresponding to $|{\bf e}_u^{uu}|$ and $|{\bf b}_u^{uu}|$, found to be $<4.308\times10^{-12}$. This method to bound SME coefficients is advantageously sensitive to effects in the second and third quark families, even though nucleon EMMs are defined solely by up- and down-quark contributions. Regarding EMMs of charged leptons, we considered the experiment--theory differences $\Delta a_A=a_A^{\rm exp}-a_A^{\rm SM}$ for the electron and muon AMMs, given in Refs.~[\refcite{electronamm,muonamm}] as $\Delta a_e=-1.06(082)\times10^{-12}$ and $\Delta a_\mu=249(87)\times10^{-11}$, respectively, and upper bounds on their EMDs, which, according to Refs.~[\refcite{electronedm,muonedm}], are $|d_e|<8.7\times10^{-29}e\cdot{\rm cm}$ and $|d_\mu|<1.8\times10^{-19}e\cdot{\rm cm}$. The electron and muon AMMs constrain SME coefficients within the range $10^{-7}$ to $10^{-11}$, with the most restrictive limits set on $|{\rm Re}\,{\bf e}_e^{ee}|$, $|{\rm Re}\,{\bf b}_e^{ee}|$, constrained to be $<1.619\times10^{-11}$, as well as $|{\rm Im}\,{\bf b}_e^{ee}|$ and $|{\rm Re}\,{\bf b}_e^{ee}|$, restricted to be $<3.620\times10^{-11}$. We use experimental limits on the electron and muon EDMs to constrain SME coefficients,\cite{AMNST} finding values within $10^{-5}$ to $10^{-14}$. The best limits are set on $|{\bf e}_e^{ee}|$ and $|{\bf b}_e^{ee}|$, both restricted to be  $<5.282\times10^{-14}$.

\section*{Acknowledgments}
The authors acknowledge financial support from CONACYT (M\'exico).

\end{document}